\documentclass[11pt]{article}
\usepackage{latexsym}
\usepackage{amssymb}
\usepackage[dvips]{graphicx}
\usepackage{epsfig}
\usepackage{rotating}
\usepackage{amsfonts}
\usepackage{amsmath}
\begin{document}

\begin{titlepage}
\begin{flushright}
CP3-19-31\\
\end{flushright}

\vspace{5pt}

\begin{center}

{\Large\bf Non Local Global Symmetries}\\

\vspace{7pt}

{\Large\bf of a Free Scalar Field}\\

\vspace{7pt}

{\Large\bf in a Bounded Spatial Domain}\\

\vspace{60pt}

Daddy Balondo Iyela$^{a,b}$ and
Jan Govaerts$^{b,c,}$\footnote{Fellow of the Stellenbosch Institute for Advanced Study (STIAS), Stellenbosch,
Republic of South Africa}

\vspace{10pt}

$^{a}${\sl D\'epartement de Physique, Universit\'e de Kinshasa (UNIKIN),\\
B.P. 190 Kinshasa XI, Democratic Republic of Congo}\\
E-mail: {\em balondo36@gmail.com}\\
\vspace{10pt}
$^{b}${\sl Centre for Cosmology, Particle Physics and Phenomenology (CP3),\\
Institut de Recherche en Math\'ematique et Physique (IRMP),\\
Universit\'e catholique de Louvain (UCLouvain),\\
2, Chemin du Cyclotron, B-1348 Louvain-la-Neuve, Belgium}\\
E-mail: {\em Jan.Govaerts@uclouvain.be}\\
\vspace{10pt}
$^{c}${\sl International Chair in Mathematical Physics and Applications (ICMPA--UNESCO Chair)\\
University of Abomey-Calavi, 072 B.P. 50, Cotonou, Republic of Benin}\\

\vspace{10pt}

%\today

\vspace{10pt}

\begin{abstract}
\noindent
BMS symmetries have been attracting a great deal of interest in recent years.
Originally discovered as being the symmetries of asymptotically flat spacetime geometries at null infinity
in General Relativity, BMS symmetries have also been shown to exist for free field theories
over Minkowski spacetime. In wanting to better understand their status and the underlying reasons
for their existence, this work proposes a general rationale towards identifying all possible global symmetries
of a free field theory over Minkowski spacetime, by allowing the corresponding conserved generators
not to be necessarily spatially local in phase space since fields and their conjugate momenta
are intrinsically spatially non local physical entities.
As a preliminary towards a separate study of the role of asymptotic states for BMS symmetries
in an unbounded Minkowski spacetime, the present discussion focuses first onto a 2+1 dimensional
free scalar field theory in a bounded spatial domain with the topology of a disk and an arbitrary radial
Robin boundary condition. The complete set of global symmetries of that system, most of which are 
dynamical symmetries but include as well those generated by the local total energy and angular-momentum
of the field, is thereby identified.

\end{abstract}

\end{center}

\end{titlepage}

\setcounter{footnote}{0}

\section{Introduction}
\label{Intro}

Consider a free field, whether classical or quantum, over a background spacetime of specific geometry,
say that of a Minkowski spacetime. Besides the symmetries of that geometry --  thus those of the Poincar\'e group
in the instance of a Minkowski spacetime -- what are all the possible global symmetries of such a field dynamics?

Of course these symmetries include the global symmetries of the background spacetime geometry.
However there are far more symmetries possible still, certainly in the case of a free field.
As is well known -- a result which came as a surprise when it was first
discovered -- in Einstein's General Relativity an asymptotically flat spacetime possesses asymptotic global symmetries at null infinity
that include the Poincar\'e group but are in fact much larger with an infinite (countable) set of conserved quantities,
namely the generators of the so-called BMS symmetries
of super-translations and super-rotations\cite{BMS1, BMS2, BMS3,Hawking,Strominger,Compere}. More recently
it has been established that free fields in a Minkowski spacetime share the same BMS symmetries
as well\cite{Free1,Free2,Free3}.
In particular in Refs.\cite{Free2,Free3} it was shown that these BMS symmetries are related
to conserved charges and generators which are bilinear in the phase space fields but are, as a matter of fact,
spatially non local.

Because of their wave-like dynamics Lorentz covariant fields are intrinsically spatially non local yet with a causally
consistent time evolution.
On the other hand, by definition symmetries of a dynamical system are transformations of its degrees
of freedom that leave its equations of motion form-invariant, namely that map solutions into solutions
to the same equations of motion. However dynamical systems may possess symmetries larger than those
that leave their Lagrangian action invariant (up to total time derivatives), namely symmetry transformations which,
when acting on the associated phase space (rather than on the system's configuration and
conjugate momentum spaces separately), leave the system's Hamiltonian action invariant (up to total time derivatives).
Such symmetry transformations mixing configuration space degrees of freedom with their conjugate momenta
correspond generally to so-called dynamical symmetries\footnote{Such as the celebrated SO(4) symmetry of the
bound states of Coulomb--Kepler problem, or the SU(3) symmetry of the spherically symmetric harmonic oscillator,
whether classical or quantum mechanical in 3 dimensional Euclidean geometry.}.
Obviously Noether's (first) theorem (for global symmetries) applies likewise
to the phase space parametrisation of a dynamics in terms of its Hamiltonian action, allowing for the identification
of the Noether conserved charges generating global symmetries over phase space\cite{Canonical}
through Poisson brackets. Consequently the general setting appropriate to studying the global symmetries
of a system is that of its canonical Hamiltonian formulation.

In the case of a free field, say a bosonic one, its action being quadratic and its equation of motion being
linear in the field, hence also its Hamiltonian equations of motion, the general transformations of the phase space
fields leaving these equations invariant, hence also the Hamiltonian action, are {\sl a priori} linear in the
phase space fields. The corresponding (Noether charge) generators thus need to be bilinear in these fields.
However because of the intrinsic spatially non local character of field dynamics, these generators need not
necessarily be local in space (as are for instance the total energy, momentum and angular-momentum
of the field when conserved), and may indeed be spatially bilocal, namely be given by a double spatial integral
of a bilinear form in the phase space fields.

This is thus the rationale to be implemented in the present paper when wanting to identify all the possible
global symmetries of a free field dynamics. Specifically the analysis to be presented applies to the
simplest case possible, namely that of a free and real bosonic scalar field in a 2+1 dimensional Minkowski space-time
(generalisation to Minkowski spacetimes of arbitrary dimension is of course feasible).
Furthermore in order to better understand the possible role of asymptotic behaviour at infinity
for the existence of BMS symmetries (in the spirit of their original discovery), in the present work
a spatially bounded domain will first\footnote{In a forthcoming paper we plan to present
a similar analysis for the free real scalar field in the full unbounded 2+1 dimensional Minkowski spacetime,
in which case the connection with the known BMS symmetries of that system will be established.} be considered,
namely that of a disk of finite radius $R>0$, $D(R)$, in order to at least preserve -- besides the invariance
under global time translations (thus the conservation of the field's total energy) -- the invariance under
spatial rotations around a fixed center as well (and thus the conservation of the field's total angular-momentum).
In other words the considered 2+1 dimensional spacetime has the topology of a plain cylinder
$\mathbb{R}\times D(R)$.
However the spatial boundary condition to be imposed on the scalar field will be as general as possible
in a manner consistent with rotational invariance, namely in terms of a radial Robin boundary condition
of uniform Robin parameter along the disk boundary, allowing for a smooth transition between a Dirichlet
and a Neumann boundary condition if required. This choice of boundary condition is to be implemented
in the field dynamics through a specific boundary term in the Lagrangian action of the free field.

With the purpose as well of specifying notations and establishing results for later use,
Section 2 discusses the free field dynamics based on an action principle, its Hamiltonian formulation and
eventually its canonical quantisation. Section 3 then implements the rationale outlined above,
to identify the complete ensemble of global symmetries of the considered free field dynamics,
specifically for the flat disk spatial topology and geometry. The comprehensive identification of the
complete global symmetry group is achieved in Section 4. While some final comments are
presented in the Conclusion.

\section{Free Field Dynamics and its Quantisation}
\label{Sect2}

Given the background 2+1 dimensional
Minkowski spacetime of topology $\mathbb{R}\times D(R)$, in natural units ($c=1=\hbar$) let us thus consider
a real scalar field $\phi(t,r,\theta)$, $(r,\theta)$ being polar coordinates in the disk ($0\le r\le R$ and $0\le\theta<2\pi$).
The dynamics of that field derives from the following choice of Lagrangian action which is purely quadratic in the field,
\begin{eqnarray}
S_0\left[\phi\right] = \int dt\,L_0 &=&\int dt\,\int_0^Rdr\,r \int_0^{2\pi}d\theta
\left[\frac{1}{2}\left(\partial_t\phi\right)^2-\frac{1}{2}\left(\partial_r\phi\right)^2
-\frac{1}{2}\left(\frac{1}{r}\partial_\theta\phi\right)^2-\frac{1}{2}\mu^2\phi^2\right]\ + \nonumber \\
&&+\int dt\,\int_0^{2\pi}d\theta\ \frac{1}{2}\lambda\ \phi^2(t,R,\theta),
\label{eq:Action1}
\end{eqnarray}
where $\mu\ge 0$ is a real mass parameter while $\lambda\in\mathbb{R}$ is the real, constant and dimensionless
Robin parameter determining the radial Robin boundary condition for the equation of motion of the free scalar field.
In the above expression the first line is the bulk contribution to the action while the second line is a boundary
contribution. Note that besides its invariances under global time translations and spatial rotations around the
center of the disk $D(R)$, this dynamics is also invariant under parity and time reversal transformations.

\subsection{Classical Solutions}
\label{Sect2.1}

Obviously given the above action the variational principle, which requires the action to be stationary only up
to a total time derivative, implies the usual Klein-Gordon wave equation for the scalar field in the disk,
but subjected as well to the radial Robin boundary condition of parameter~$\lambda$,
\begin{equation}
\left(\partial^2_t-\frac{1}{r}\partial_r\,r\partial_r -\frac{1}{r^2}\partial^2_\theta + \mu^2\right)\phi(t,r,\theta)=0,\qquad
\left(\partial_r\phi\right)(t,R,\theta)=\frac{\lambda}{R}\,\phi(t,R,\theta).
\end{equation}
Note that the value $\lambda=0$ reproduces a Neumann boundary condition,
while the limits $\lambda\rightarrow\pm\infty$ imply a Dirichlet boundary condition for $\phi(t,r,\theta)$.

The Klein-Gordon equation being linear in the field, through a separation of variables
it suffices to find a basis of solutions to construct its general solution by superposition.
The lack of spatial translational invariance in the present instance however, implies that the usual
travelling plane waves are not available for such a basis. A basis of solutions needs rather to be
identified in terms of stationary waves. Given the invariances of the dynamics under global
translations in time and global rotations in the disk with a $2\pi$ periodicity in the $\theta$ dependence,
through a separation of variables such stationary waves of angular-momentum $\ell\in\mathbb{Z}$
and positive angular frequency $\omega_\ell>0$ may be considered in the following (complex) form,
\begin{equation}
\phi_\ell(t,r,\theta)=e^{-i\omega_\ell t}\,e^{i\ell\theta}\,f_\ell(r),\qquad
\omega_\ell>0,\qquad \ell\in\mathbb{Z}.
\end{equation}
Direct substitution into the Klein-Gordon equation then requires,
\begin{equation}
\left[\frac{d^2}{dx^2}+\frac{1}{x}\frac{d}{dx}+\left(1-\frac{\ell^2}{x^2}\right)\right]\,f_\ell(r)=0,\qquad x=k_\ell\, r,
\qquad k_\ell>0,
\end{equation}
where $\omega^2_\ell=k^2_\ell+\mu^2\ge \mu^2$, $k_\ell \equiv \sqrt{\omega^2_\ell-\mu^2}\ge 0$,
$\omega_\ell\ge\mu\ge 0$.
This is the differential equation defining the real Bessel and Neumann functions of order $\ell$,
$J_\ell(x)$ and $N_\ell(x)$. Since the field $\phi(t,r,\theta)$ must remain regular throughout the disk,
only the Bessel functions of the first kind, $J_\ell(x)$, may be retained as solution for the radial dependence
of the field, namely $f_\ell(r)=J_\ell(k_\ell r)$ up to a constant normalisation factor. Let us then also recall
the parity property $J_{-\ell}(x)=(-1)^\ell\,J_\ell(x)$ for $\ell\in\mathbb{Z}$, hence equivalently
$i^{-\ell} J_{-\ell}(x)=i^\ell J_\ell(x)$.

However the spectrum of wave number values $k_\ell$ is restricted by imposing the Robin boundary condition,
which leads to the following equation in $k_\ell$ ($J'_\ell(x)$ denotes the derivative of $J_\ell(x)$ relative
to its argument $x$),
\begin{equation}
x_R\,J'_\ell(x_R)\,-\,\lambda\,J_\ell(x_R)=0,\qquad x_R=k_\ell R,\qquad
k_\ell\,J'(k_\ell R)=\frac{\lambda}{R}\,J_\ell(k_\ell R).
\end{equation}
The properties of the positive roots of this equation in $x_R$ have been discussed
in detail in Refs.\cite{Roots1,Roots2}.
For generic values of the Robin parameter $\lambda$ and a given value of $\ell$,
the positive roots form an infinite discrete set of
non degenerate values, $x_{\ell,n}$, increasing with $n=1,2,\cdots$, the smallest of which is strictly positive,
$x_{\ell,1}>0$. Correspondingly the spectrum of wave numbers solving the Robin boundary condition is thus
given as $k_{\ell,n}=x_{\ell,n}/R$, with $\omega_{\ell,n}=\sqrt{k^2_{\ell,n}+\mu^2}$.
As a function of $R$ and for a given value of $\ell$,
it is only at most for a finite set of isolated values for $\lambda$ (if not a single such value)
that the smallest root, $x_{\ell,1}$, may be doubly degenerate. In this work we shall assume that
such a degenerate circumstance is not encountered nor that accidental degeneracies in the roots $x_{\ell,n}$
would occur for different values of $|\ell |$ (in which case otherwise the set of global
symmetries to be identified for the free field theory would become slightly enlarged as compared
to our final identification of the global symmetries of this system).

However, note that because of parity invariance, or equivalently time reversal invariance of the field,
these roots are doubly degenerate for opposite non vanishing values of $\ell$,
\begin{equation}
x_{-\ell,n}=x_{\ell,n},\quad
k_{-\ell,n}=k_{\ell,n},\qquad
\omega_{-\ell,n}=\omega_{\ell,n},\qquad \ell\ne 0,\quad \ell\in\mathbb{Z},
\end{equation}
as also follows form the fact that $J_{-\ell}(x)=(-1)^\ell J_\ell(x)$.

In conclusion the full spectrum of solutions to the Laplacian eigenvalue problem on the disk,
$\Delta\varphi(r,\theta)=-k^2\varphi(r,\theta)$ with $k\in\mathbb{R}$ and $k\ge 0$,
subjected to the rotationally invariant radial Robin boundary condition of parameter $\lambda$, namely,
\begin{equation}
\left(\frac{1}{r}\partial_r r\partial_r + \frac{1}{r^2}\partial^2_\theta\right)\varphi(r,\theta)=-k^2\,\varphi(r,\theta),\qquad
(\partial_r\varphi)(R,\theta)=\lambda\,\varphi(R,\theta),
\end{equation}
is given by the following basis of functions, with $x_{\ell,n}=k_{\ell,n}R$,
\begin{equation}
\varphi_{\ell,n}(r,\theta) = N_{\ell,n}\,e^{i\ell\theta}\,J_\ell\left(x_{\ell,n}\frac{r}{R}\right),\qquad
k_{\ell,n}\,J'_\ell(k_{\ell,n}R)=\frac{\lambda}{R}\,J_\ell(k_{\ell,n}R),\qquad x_{\ell,n}=k_{\ell,n}R,
\end{equation}
where $N_{\ell,n}$ are (possibly complex) constant normalisation factors still to be determined.
Requiring these functions to be orthonormalised in the disk, namely
\begin{equation}
\int_0^Rdr\, r\int_0^{2\pi}d\theta\,\varphi^*_{\ell_1,n_1}(r,\theta)\,\varphi_{\ell_2,n_2}(r,\theta)
=\delta_{\ell_1,\ell_2}\,\delta_{n_1,n_2},
\end{equation}
implies for the normalisation factors\footnote{Using the differential equation obeyed by $J_n(kx)$ (with
$n\in\mathbb{N}$)
it is possible to establish that $\int dx \, x J_n(kx)\,J_n(\ell x)=\frac{x}{k^2-\ell^2}\left(J_n(kx)\frac{dJ_n(\ell x)}{dx}
-\frac{dJ_n(k x)}{dx}J_n(\ell x)\right)$ for $k^2\ne \ell^2$, as well as
$\int dx\,x J^2_n(\ell x)=\frac{1}{2\ell^2}\left[\left(x\frac{dJ_n(\ell x)}{dx}\right)^2 + (\ell^2 x^2 - n^2) J^2_n(\ell x)\right]$.
Orthogonality of the functions $\varphi_{\ell,n}(r,\theta)$ for different values of $n$ (and a same value for $\ell$)
then follows by relying on the Robin boundary condition identity defining the roots $x_{\ell,n}$.},
\begin{equation}
{\cal N}_{\ell,n}\equiv
|N_{\ell,n}|=\left(\frac{\pi R^2}{x^2_{\ell,n}}\left(\lambda^2+x^2_{\ell,n}-\ell^2\right)\,J^2_\ell(x_{\ell,n})\right)^{-1/2}
=\frac{k_{\ell,n}}{\sqrt{\pi\left(\lambda^2+x^2_{\ell,n}-\ell^2\right)\,J^2_\ell(x_{\ell,n})}}.
\end{equation}
Finally because of the well known representation
\begin{equation}
e^{ix\cos\theta}=\sum_{\ell=-\infty}^\infty\,i^\ell\,e^{i\ell\theta}\,J_\ell(x),
\end{equation}
our choice of phase convention for the normalisation factors $N_{\ell,n}$ is such that
\begin{equation}
\varphi_{\ell,n}(r,\theta)={\cal N}_{\ell,n}\,i^\ell\,e^{i\ell\theta}\,J_\ell(k_{\ell,n}r).
\end{equation}
Besides the above orthonormality properties the functions $\varphi_{\ell,n}(r,\theta)$ also possess the following
completeness relation as a genuine basis of functions in polar coordinates on the disk
obeying the radial Robin boundary condition of parameter $\lambda$,
\begin{equation}
\sum_{\ell=-\infty}^\infty\,\sum_{n=1}^\infty\,\varphi_{\ell,n}(r_1,\theta_1)\,\varphi^*_{\ell,n}(r_2,\theta_2)=
\frac{1}{r_1}\,\delta(r_1 - r_2)\,\delta(\theta_1 - \theta_2).
\label{eq:complete}
\end{equation}
Note that because of the properties under the change of sign in $\ell$ pointed out above,
and the phase convention for $N_{\ell,n}$ specified above, we also have,
\begin{equation}
\varphi_{-\ell,n}(r,\theta)=(-1)^\ell\,\varphi^*_{\ell,n}(r,\theta),
\end{equation}
in a way similar to an analoguous property obeyed by the usual spherical harmonics on the 2-sphere.

In conclusion the complete and general solution to the Klein-Gordon equation subjected to the considered
radial Robin boundary condition is thus of the form,
\begin{equation}
\phi(t,r,\theta)=\sum_{\ell=-\infty}^\infty\sum_{n=1}^\infty\frac{1}{\sqrt{2\omega_{\ell,n}}}
\left[e^{-i\omega_{\ell,n}t}\,\varphi_{\ell,n}(r,\theta)\,a(\ell,n)\,+\,
e^{+i\omega_{\ell,n} t}\,\varphi^*_{\ell,n}(r,\theta)\,a^\dagger(\ell,n)\right],
\label{eq:sol1}
\end{equation}
where the choice of normalisation is made for later convenience, while $a^\dagger(\ell,n)\equiv a^*(\ell,n)$
and $a(\ell,n)$ are complex valued integration constants corresponding to the excitation amplitudes and phases
of each of the possible excitation modes $(\ell,n)$ of standing waves in the disk, which for the quantised
field will correspond to the (normalised) Fock algebra creation and annihilation operators for the different quanta
possible for this free field theory.

Within the Hamiltonian formulation of the system (see the next Subsection),
in polar coordinates in the disk the momentum conjugate to the field, $\pi(t,r,\theta)$, is given
as $\pi(t,r,\theta)=r\partial_t\phi(t,r,\theta)$, namely,
\begin{equation}
\pi(t,r,\theta)=\sum_{\ell=-\infty}^\infty\sum_{n=1}^\infty\frac{-i\omega_{\ell,n}}{\sqrt{2\omega_{\ell,n}}}
\left[e^{-i\omega_{\ell,n}t}\,r\varphi_{\ell,n}(r,\theta)\,a(\ell,n)\,-\,
e^{+i\omega_{\ell,n} t}\,r\varphi^*_{\ell,n}(r,\theta)\,a^\dagger(\ell,n)\right].
\label{eq:sol2}
\end{equation}
The relevant inverse relations are then, for the corresponding mode amplitudes,
\begin{equation}
a(\ell,n)=\int_0^Rdr\,r\int_0^{2\pi}d\theta\frac{\sqrt{2\omega_{\ell,n}}}{2}\,e^{i\omega_{\ell,n}t}\,\varphi^*_{\ell,n}(r,\theta)
\left[\phi(t,r,\theta) + \frac{i}{\omega_{\ell,n}}\frac{1}{r}\pi(t,r,\theta)\right],
\label{eq:a}
\end{equation}
\begin{equation}
a^\dagger(\ell,n)=\int_0^Rdr\,r\int_0^{2\pi}d\theta\frac{\sqrt{2\omega_{\ell,n}}}{2}
\,e^{-i\omega_{\ell,n}t}\,\varphi_{\ell,n}(r,\theta)
\left[\phi(t,r,\theta) - \frac{i}{\omega_{\ell,n}}\frac{1}{r}\pi(t,r,\theta)\right].
\label{eq:adagger}
\end{equation}

\subsection{Hamiltonian Formulation}
\label{Sect2.2}

Given the polar coordinate parametrisation of the Lagrangian action in (\ref{eq:Action1}) the momentum conjugate
to the field configuration $\phi(t,r,\theta)$, defined by $\pi(t,r,\theta)=\delta L/\delta(\partial_t\phi(t,r,\theta))$,
is obtained as,
\begin{equation}
\pi(t,r,\theta)=r\,\partial_t\phi(t,r,\theta),\qquad
\partial_t\phi(t,r,\theta)=\frac{1}{r}\pi(t,r,\theta),
\end{equation}
for which the Poisson brackets are canonical, with the only non vanishing (equal time) bracket being
\begin{equation}
\{\phi(t,r_1,\theta_1),\pi(t,r_2,\theta_2)\}=\delta(r_1-r_2) \delta(\theta_1 - \theta_2).
\end{equation}
Correspondingly the canonical Hamiltonian of the system is given as, inclusive of the boundary term,
\begin{equation}
H=\int_0^R dr\,\int_0^{2\pi}d\theta
\left(\frac{1}{2r}\pi^2 + \frac{1}{2}r(\partial_r\phi)^2+\frac{1}{2r}(\partial_\theta\phi)^2+\frac{1}{2}\mu^2 r\phi^2\right)
\ +\ \int_0^{2\pi}d\theta\left(-\frac{1}{2}\lambda\,\phi^2(t,R,\theta)\right),
\label{eq:H}
\end{equation}
in terms of which the Hamiltonian first-order action is expressed as
\begin{equation}
S_H[\phi,\pi]=\int dt\,L_H = \int dt\,\left[\int_0^R dr\, \int_0^{2\pi}d\theta\,\partial_t \phi \, \pi\,-\,H\right].
\label{eq:SH}
\end{equation}
It should be noted that in view of its definition the conjugate momentum field obeys a rotationally invariant
radial Robin boundary condition of parameter $(\lambda+1)$, namely,
\begin{equation}
(\partial_r\phi)(t,R,\theta)=\frac{\lambda}{R}\,\phi(t,R,\theta),\qquad
(\partial_r\pi)(t,R,\theta)=\frac{\lambda+1}{R}\,\pi(t,R,\theta).
\end{equation}
These are thus the boundary conditions to be considered when solving the associated Hamiltonian equations
of motion, of which the solutions will be expressed in terms of the basis of functions $\varphi_{\ell,n}(r,\theta)$ and
$r\varphi_{\ell,n}(r,\theta)$ (see (\ref{eq:sol1}) and (\ref{eq:sol2})). Note that given the equation obeyed by the roots $x_{\ell,n}=k_{\ell,n}R$, we have indeed as it should,
\begin{equation}
\frac{d}{dr}J_\ell(k_{\ell,n}r)_{|_{r=R}}=\frac{\lambda}{R}\,J_\ell(k_{\ell,n}r)_{|_{r=R}},\qquad
\frac{d}{dr}\left(rJ_\ell(k_{\ell,n}r)\right)_{|_{r=R}}=\frac{\lambda+1}{R}\left(rJ_\ell(k_{\ell,n}r)\right)_{|_{r=R}}.
\end{equation}

Through a careful treatment of the boundary term contribution in the evaluation of Poisson brackets with
the above Hamiltonian $H$, it may readily be checked that the first-order and linear Hamiltonian equations
of motion read as
\begin{equation}
\partial_t \phi=\left\{\phi,H\right\}=\frac{1}{r}\pi,\qquad
\partial_t\pi=\left\{\pi,H\right\}=r\left(\frac{1}{r}\partial_r r\partial_r + \frac{1}{r^2}\partial^2_\theta - \mu^2\right)\phi
=r\left(\Delta - \mu^2\right)\phi,
\end{equation}
of which, given the appropriate Robin boundary conditions, the solutions are provided by the expressions
(\ref{eq:sol1}) and (\ref{eq:sol2}). Note that as it should, these same equations of motion also derive directly
from the variational principle applied to the above Hamiltonian action $S_H$, inclusive of spatial boundary
contributions that need to vanish as well.

Using the inverse relations in (\ref{eq:a}) and (\ref{eq:adagger}) as well as the orthonormality properties
of the basis functions $\varphi_{\ell,n}(r,\theta)$, one may also readily establish that the above canonical
Poisson brackets for the phase space fields $\phi(t,r,\theta)$ and $\pi(t,r,\theta)$ translate into
the following (only non vanishing) Poisson brackets for the mode amplitudes,
\begin{equation}
\left\{a(\ell_1,n_2),a^\dagger(\ell_2,n_2)\right\}=-i\,\delta_{\ell_1,\ell_2}\,\delta_{n_1,n_2}.
\end{equation}

Given the expression (\ref{eq:H}) and using the Robin boundary condition for $\phi$, through partial integrations
it is straightforward to establish the following purely bulk representation for the Hamiltonian which is spatially
local in phase space and quadratic in the phase space fields,
\begin{equation}
H=\int_0^R dr\, r \int_0^{2\pi}d\theta\left[\frac{1}{2}\left(\frac{1}{r}\pi\right)^2
-\frac{1}{2}\phi\left(\frac{1}{r}\partial_r r\partial_r + \frac{1}{r^2}\partial^2_\theta - \mu^2\right)\phi\right].
\label{eq:H2}
\end{equation}
Then using the fact that,
\begin{equation}
\left(\Delta - \mu^2\right)\varphi_{\ell,n}(r,\theta)=
\left(\frac{1}{r}\partial_r r\partial_r + \frac{1}{r^2}\partial^2_\theta - \mu^2\right)\varphi_{\ell,n}(r,\theta)
=-\omega^2_{\ell,n}\,\varphi_{\ell,n}(r,\theta),\quad
\omega^2_{\ell,n}=k^2_{\ell,n}+\mu^2,
\end{equation}
as well as the orthonormality properties of the basis functions $\varphi_{\ell,n}(r,\theta)$,
a straightforward evaluation establishes the following representation for the conserved total energy of the field,
in which none of the factors $a(\ell,n)$ and $a^\dagger(\ell,n)$ have been permuted with one another.
\begin{equation}
H=\sum_{\ell=-\infty}^\infty\sum_{n=1}^\infty\frac{1}{2}\omega_{\ell,n}
\left(a^\dagger(\ell,n) a(\ell,n) + a(\ell,n) a^\dagger(\ell,n)\right).
\end{equation}

Besides the global symmetry under constant translations in time for which the conserved Noether charge
is the total energy or the Hamiltonian $H$, the system also possess a global symmetry under constant
rotations of the disk around its center at $r=0$, for which the conserved Noether charge is
the total angular-momentum of the field, $L$. A simple analysis identifies its spatially local phase space
expression which is quadratic in the phase space fields,
\begin{equation}
L=-\int_0^R dr\, r \int_0^{2\pi}d\theta\,\left(\frac{1}{r}\pi\right)\,\partial_\theta\phi.
\end{equation}
A straighforward substitution in terms of the mode amplitudes then finds,
\begin{equation}
L=\sum_{\ell=-\infty}^\infty\sum_{n=1}^\infty\,
\frac{1}{2}\ell\left(a^\dagger(\ell,n) a(\ell,n) + a(\ell,n) a^\dagger(\ell,n)\right).
\end{equation}

\subsection{The Quantised Field}
\label{Sect2.3}

In terms of the results established above within the Hamiltonian formulation of the system,
its canonical quantisation is straightforward. Its Hilbert space of quantum states is spanned
by the tensor product of the Fock spaces generated from a Fock vacuum by the Fock algebras
of annihilation and creation operators, $a(\ell,n)$ and $a^\dagger(\ell,n)$, obeying the Fock algebras
(with $\hbar=1$),
\begin{equation}
\left[a(\ell_1,n_1),a^\dagger(\ell_2,n_2)\right]=\delta_{\ell_1,\ell_2}\,\delta_{n_1,n_2}\,\mathbb{I},
\qquad \ell_1,\ell_2\in\mathbb{Z},\quad n=1,2,3,\cdots .
\end{equation}
Time dependence of the system is generated through the total quantum Hamiltonian, which we choose to be given
by the normal ordered form of the classical quantity, namely,
\begin{equation}
\hat{H}=\sum_{\ell=-\infty}^\infty\sum_{n=1}^\infty\,\omega_{\ell,n}\,a^\dagger(\ell,n) a(\ell,n).
\end{equation}
In particular in the Heisenberg picture the time dependent
quantum field operators $\hat{\phi}(t,r,\theta)$ and $\hat{\pi}(t,r,\theta)$
remain then given by the same expressions as in (\ref{eq:sol1}) and (\ref{eq:sol2}),
now in terms of the Fock operators.

Likewise the total angular-momentum operator of the system is represented by the normal ordered expression,
\begin{equation}
\hat{L}=\sum_{\ell=-\infty}^\infty\sum_{n=1}^\infty\,\ell\,a^\dagger(\ell,n) a(\ell,n).
\end{equation}

Hence the action of $a^\dagger(\ell,n)$ onto any quantum state of the system adds one more quantum
of energy $\omega_{\ell,n}$ and angular-momentum $\ell$
in the standing wave mode of the field of quantum numbers $(\ell,n)$
related to the root $x_{\ell,n}$ of the equation expressing the radial Robin boundary condition
for the angular-momentum mode $\ell$.
Because of the parity and time reversal invariance property that $x_{-\ell,n}=x_{\ell,n}$,
note well that the energy spectrum of the system is doubly degenerate under the
exchange $\ell\leftrightarrow -\ell$ for $\ell\ne 0$. In other words one may as well write,
\begin{equation}
\hat{H}=\sum_{n=1}^\infty \omega_{0,n}\, a^\dagger(0,n)a(0,n)\,+\,
\sum_{\ell=1}^\infty\sum_{n=1}^\infty\omega_{\ell,n}\left(a^\dagger(\ell,n)a(\ell,n)+a^\dagger(-\ell,n)a(-\ell,n)\right),
\label{eq:Hsu2}
\end{equation}
\begin{equation}
\hat{L}= \sum_{\ell=1}^\infty\sum_{n=1}^\infty \ell \left(a^\dagger(\ell,n)a(\ell,n)-a^\dagger(-\ell,n)a(-\ell,n)\right).
\label{eq:Lsu2}
\end{equation}

\section{Bilocal Global Dynamical Symmetries and their Charges}
\label{Sect3}

Given the rationale outlined in the Introduction, let us now consider the possibility that quantities
which are bilocal and bilinear in the phase space fields, hence bilinear in the creation and annihilation
operators, could generate (as the would-be associated conserved Noether charges) global symmetries
of the free field dynamics. In a compact notation such that ``1" or ``2" stand for the pairs of coordinates
$(r_1,\theta_1)$ and $(r_2,\theta_2)$, respectively, such candidate conserved charges are of the general form
\begin{eqnarray}
Q(t) &=& \int_0^R dr_1\int_0^{2\pi}d\theta_1\,\int_0^R dr_2\int_0^{2\pi}d\theta_2 \times \nonumber \\
&&\times \left(\frac{1}{2}f(1;2)\phi(t,1)\phi(t,2)+\frac{1}{2}g(1;2)\pi(t,1)\pi(t,2)+h(1;2)\phi(t,1)\pi(t,2)\right),
\end{eqnarray}
where the triplet of time independent functions $f(1;2)\equiv f(r_1,\theta_1;r_2,\theta_2)$,
$g(1;2)\equiv g(r_1,\theta_1;r_2,\theta_2)$ and $h(1;2)\equiv h(r_1,\theta_1;r_2,\theta_2)$
are real functions to be identified. In particular note that
the functions $f(1;2)$ and $g(1;2)$ are symmetric under the exchange $1\leftrightarrow 2$, while
$h(1;2)$ does not possess any specific symmetry under that exchange,
\begin{equation}
f(2;1)=f(1;2),\qquad
g(2;1)=g(1;2).
\end{equation}

Through Poisson brackets with the fields $\phi(t,r,\theta)$ and $\pi(t,r,\theta)$ such charges $Q(t)$
generate the following linear transformations mixing the phase space fields,
\begin{equation}
\delta \phi(t,r,\theta)=\int_0^R dr_1\int_0^{2\pi}d\theta_1
\left[g(r,\theta;r_1,\theta_1)\,\pi(t,r_1,\theta_1)\,+\,h(r_1,\theta_1;r,\theta)\,\phi(t,r_1,\theta_1)\right],
\label{eq:var1}
\end{equation}
\begin{equation}
\delta \pi(t,r,\theta)=-\int_0^R dr_1\int_0^{2\pi}d\theta_1
\left[f(r,\theta;r_1,\theta_1)\,\phi(t,r_1,\theta_1)\,+\,h(r,\theta;r_1,\theta_1)\,\pi(t,r_1,\theta_1)\right].
\label{eq:var2}
\end{equation}
In a form more convenient for later use, these transformations may also be expressed as,
\begin{equation}
\delta \phi(t,r,\theta)=\int_0^R dr_1\,r_1\int_0^{2\pi}d\theta_1
\left[g(r,\theta;r_1,\theta_1)\,\left(\frac{1}{r_1}\pi(t,r_1,\theta_1)\right)
\,+\,\frac{1}{r_1}h(r_1,\theta_1;r,\theta)\,\phi(t,r_1,\theta_1)\right],
\label{eq:var3}
\end{equation}
\begin{eqnarray}
\delta\left(\frac{1}{r}\pi(t,r,\theta)\right) &=& -\int_0^R dr_1\,r_1\int_0^{2\pi}d\theta_1 \times \nonumber \\
&&\ \times\left[ \frac{1}{r r_1}f(r,\theta;r_1,\theta_1)\,\phi(t,r_1,\theta_1)\,+\,
\frac{1}{r}h(r,\theta;r_1,\theta_1)\,\left(\frac{1}{r_1}\pi(t,r_1,\theta_1)\right)\right].
\label{eq:var4}
\end{eqnarray}

A first series of restrictions on the functions $f$, $g$ and $h$ is implied by the necessity that the transformed
fields still obey their relevant Robin boundary conditions of parameter $\lambda$ or $(\lambda+1)$ as the case
may. Consequently one must require,
\begin{equation}
\partial_r g(r,\theta;r_1,\theta_1)_{|_{r=R}}=\frac{\lambda}{R}\ g(R,\theta;r_1,\theta_1),\qquad
\partial_r h(r_1,\theta_1;r,\theta)_{|_{r=R}}=\frac{\lambda}{R}\ h(r_1,\theta_1;R,\theta),
\end{equation}
\begin{equation}
\partial_r f(r,\theta;r_1,\theta_1)_{|_{r=R}}=\frac{\lambda+1}{R}\ f(R,\theta;r_1,\theta_1),\qquad
\partial_r h(r,\theta;r_1,\theta_1)_{|_{r=R}}=\frac{\lambda+1}{R}\ h(R,\theta;r_1,\theta_1).
\end{equation}
Note that since $\phi(t,r,\theta)$ and $\pi(t,r,\theta)/r$ both obey the radial Robin boundary condition
with a same parameter $\lambda$, as a matter of fact so do the functions $f(1;2)/(r_1 r_2)$,
$g(1;2)$ and $h(1;2)/r_1$, which is equivalent to the above boundary conditions
for the triplet of functions $(f,g,h)$.

Different strategies may be considered in order to determine under which conditions the charges $Q(t)$
generate global symmetries of the field dynamics. By definition a symmetry leaves the equations of motion
form-invariant; through a direct substitution into the Hamiltonian equations of motion of the above
transformations $\delta\phi$ and $\delta\pi$ one may find specific restrictions that the functions $f$, $g$ and $h$
would have to meet. However there would then still remain the necessity to check under which conditions
the composition (through Poisson brackets or commutators) of transformations generated by such charges $Q(t)$
for different such triples of functions $(f,g,h)$ closes on the same class of transformations.

A rather more efficient approach considers directly the Hamiltonian action of the system, $S_H$.
If the above transformations define a global symmetry leaving the equations of motion form-invariant,
necessarily they leave the action invariant possibly up to a total time derivative. If a total time derivative
is indeed thereby induced it could be that the associated Noether charge, even though conserved under its
complete equation of motion\footnote{An explicitly time dependent phase space observable, $Q(t)$,
may be a conserved quantity nevertheless, provided its Poisson bracket (or quantum commutator)
with the Hamiltonian is non vanishing and such that $dQ(t)/dt=\partial Q(t)/\partial t+\left\{Q(t),H\right\}=0$,
namely $\left\{Q(t),H\right\}=-\partial Q(t)/\partial t\ne 0$.}, possesses an explicit time dependence as a function
over phase space\cite{Canonical}, which in turn may lead to a classical central extension of the algebra spanned
by the Noether generators of such global symmetries of the system\footnote{Incidentally, on account of
Noether's (first) theorem a (global) continuous symmetry generator is always an (on-shell) conserved quantity
over phase-space. The converse however, is not necessarily true: a conserved phase space quantity is
not necessarily the generator of a symmetry. We shall see an explicit example later on.}.

Let us thus consider the Hamiltonian action (\ref{eq:SH}) with the Hamiltonian given as in (\ref{eq:H2}) (without
a boundary term), and determine its linearised variation under the transformations $\delta\phi$ and $\delta\pi$
in (\ref{eq:var1}) and (\ref{eq:var2}). Through a series of careful integrations by parts in $r$ and in $\theta$, 
and by exploiting the Robin boundary conditions that the triplet of functions $(f,g,h)$ must satisfy,
the linearised variation of the Hamiltonian action reduces to the expression,
\begin{eqnarray}
\delta S_H &=& \int dt \int_0^R dr_1\int_0^{2\pi}d\theta_1 \int_0^R dr_2\int_0^{2\pi} d\theta_2\times \nonumber \\
&\times&\Bigg\{\partial_t\left[\frac{1}{2}g(1;2)\pi(t,1)\pi(t,2)-\frac{1}{2}f(1;2)\phi(t,1)\phi(t,2)\right] + \Bigg. \nonumber \\
&&\qquad +\ \frac{1}{r_1}\pi(t,1)\pi(t,2)h(1;2) \ + \nonumber \\
&&\qquad +\ \frac{1}{r_2}\phi(t,1)\pi(t,2)f(1;2) + \phi(t,1)\pi(t,2)\,r_1(\Delta_1-\mu^2)g(1;2) \ + \nonumber \\
&&\qquad \Bigg.+\ \phi(t,1)\phi(t,2)\,r_1(\Delta_1-\mu^2) h(2;1)\Bigg\}.
\end{eqnarray}
In order that this variation reduces solely to a total time derivative contribution, in addition to those properties
already listed above the triplet of functions $(f,g,h)$ must satisfy as well the following further properties,
\begin{eqnarray}
&&\frac{1}{r_1}h(1;2)+\frac{1}{r_2}h(2;1) = 0, \nonumber \\
&&f(1;2) = -r_1 r_2(\Delta_1 -\mu^2)g(1;2), \\
&&r_1(\Delta_1 - \mu^2) h(2;1) + r_2 (\Delta_2 -\mu^2)h(1;2) = 0. \nonumber
\end{eqnarray}
Incidentally recall that $f(1;2)$ needs to be symmetric under the exchange $1\leftrightarrow 2$.
Hence we have established so far that
\begin{eqnarray}
Q(t) &=& \int_0^R dr_1\,r_1\int_0^{2\pi}d\theta_1\,\int_0^Rdr_2\,r_2\int_0^{2\pi}d\theta_2 \times
\left\{\frac{1}{2}\left(\frac{1}{r_1}\pi(t,1)\right)\left(\frac{1}{r_2}\pi(t,2)\right)\,g(1;2)\,-\, \right. \nonumber \\
&& \qquad \left. - \frac{1}{2}\phi(t,1)\phi(t,2)\,(\Delta_1-\mu^2)\,g(1;2)\,+\,
\phi(t,1)\left(\frac{1}{r_2}\pi(t,2)\right)\,\left(\frac{1}{r_1}h(1;2)\right)\right\}.
\label{eq:Q2}
\end{eqnarray}

Note that the functions $g(1;2)$, $h(1;2)/r_1$, and $f(1;2)/(r_1r_2)$, all satisfy radial Robin boundary
conditions of parameter $\lambda$, in both their dependencies in $r_1$ and $r_2$. Given that the functions
$\varphi_{\ell,n}(r,\theta)$ provide a basis of functions defined on the disk $D(R)$ and obeying this very Robin
boundary condition, it proves relevant to expand in that basis the symmetry transformation functions just listed
in order to solve the different conditions they must meet, in the following form,
\begin{equation}
g(r_1,\theta_1;r_2,\theta_2)=\sum_{\ell_1,n_1}\sum_{\ell_2,n_2}\alpha(\ell_1,n_1;\ell_2,n_2)\,
\varphi_{\ell_1,n_1}(r_1,\theta_1)\,\varphi_{\ell_2,n_2}(r_2,\theta_2),
\end{equation}
and,
\begin{equation}
\frac{1}{r_1}\,h(r_1,\theta_1;r_2,\theta_2)=\sum_{\ell_1,n_1}\sum_{\ell_2,n_2}\beta(\ell_1,n_1;\ell_2,n_2)\,
\varphi_{\ell_1,n_1}(r_1,\theta_1)\,\varphi_{\ell_2,n_2}(r_2,\theta_2),
\end{equation}
where the complex coefficients $\alpha(\ell_1,n_1;\ell_2,n_2)$ and $\beta(\ell_1,n_1;\ell_2,n_2)$
are to be restricted in order to solve for all the necessary properties that the functions $g(1;2)$ and $h(1;2)/r_1$
must possess, while each of the double sums runs over $\ell\in\mathbb{Z}$ and $n=1,2,3,\cdots$.

The function $g(1;2)$ needs to be real valued and symmetric in $(1 \leftrightarrow 2)$,
while $f(1;2)=-r_1r_2(\Delta_1 - \mu^2)g(1;2)$ needs to be real and symmetric under that exchange as well.
These three conditions translate into the following restrictions for $\alpha(\ell_1,n_1;\ell_2,n_2)$,
\begin{eqnarray}
&&\alpha^*(\ell_1,n_1;\ell_2,n_2)=(-1)^{\ell_1+\ell_2}\,\alpha(-\ell_1,n_1;-\ell_2,n_2), \nonumber \\
&&\alpha(\ell_1,n_1;\ell_2,n_2)-\alpha(\ell_2,n_2;\ell_1,n_1)=0, \\
&&\left(\omega^2_{\ell_1,n_1}-\omega^2_{\ell_2,n_2}\right)\,\alpha(\ell_1,n_1;\ell_2,n_2)=0. \nonumber
\end{eqnarray}
Given the distributions of the roots $x_{\ell,n}$ for which we assume (as a function of the value for $\lambda$)
the generic situation with only a double degeneracy under $\ell\leftrightarrow -\ell$ for fixed $n$,
the general solution to these conditions is as follows.
When $|\ell_1|\ne |\ell_2|$ all coefficients $\alpha(\ell_1,n_1;\ell_2,n_2)$ vanish identically.
When $|\ell_1|=|\ell_2|$ but $n_1\ne n_2$ again they all vanish. While for $|\ell_1|=|\ell_2|$ and $n_1=n_2$
we have the following non vanishing values, for $\ell\in\mathbb{Z}$ and $n=1,2,3,\cdots$,
\begin{eqnarray}
\alpha(\ell,n;\ell,n)\equiv\alpha_+(\ell,n) \in \mathbb{C},\qquad &{\rm with}&\quad \alpha^*_+(\ell,n)=\alpha_+(-\ell,n); \\
\alpha(\ell,n;-\ell,n)\equiv\alpha_-(\ell,n) \in\mathbb{R},\qquad &{\rm with}&\quad \alpha_-(-\ell,n)=\alpha_-(\ell,n),
\end{eqnarray}
where $\alpha_+(\ell,n)$ and $\alpha_-(\ell,n)$ are arbitrary complex and real coefficients, respectively.
Consequently one has,
\begin{equation}
g(r_1,\theta_1;r_2,\theta_2)=\sum_{\ell,n}\alpha_+(\ell,n)\,\varphi_{\ell,n}(r_1,\theta_1)\,\varphi_{\ell,n}(r_2,\theta_2)\,+\,
\sum_{\ell,n}\alpha_-(\ell,n)\,\varphi_{\ell,n}(r_1,\theta_1)\,\varphi_{-\ell,n}(r_2,\theta_2).
\end{equation}

The function $h(1;2)$ needs to be real, and to obey the two conditions
\begin{equation}
\frac{1}{r_1}\,h(1;2)\,+\,\frac{1}{r_2}\,h(2;1)=0,\qquad
\Delta_1\left(\frac{1}{r_2}\,h(2;1)\right)\,+\,\Delta_2\left(\frac{1}{r_1}\,h(1;2)\right)=0.
\end{equation}
These conditions then translate into the following restrictions for the coefficients $\beta(\ell_1,n_1;\ell_2,n_2)$,
\begin{eqnarray}
&&\beta(\ell_1,n_1;\ell_2,n_2)=(-1)^{\ell_1+\ell_2}\,\beta(-\ell_1,n_1;-\ell_2,n_2), \nonumber \\
&&\beta(\ell_1,n_1;\ell_2,n_2)\,+\,\beta(\ell_2,n_2;\ell_1,n_1)=0, \\
&&\left(k^2_{\ell_1,n_1}\,-\,k^2_{\ell_2,n_2}\right)\,\beta(\ell_1,n_1;\ell_2,n_2)=0. \nonumber
\end{eqnarray}
As a consequence all the coefficients $\beta(\ell_1,n_1;\ell_2,n_2)$ vanish identically unless
$\ell_1=\ell=-\ell_2$ and $n_1=n_2$, in which case one finds, with $\ell\in\mathbb{Z}$ and $n=1,2,3,\cdots$,
\begin{equation}
\beta(\ell,n;-\ell,n)=i\beta(\ell,n),\qquad {\rm with}\quad \beta(\ell,n)\in\mathbb{R}
\qquad {\rm and}\quad \beta(-\ell,n)=-\beta(\ell,n),
\end{equation}
where $\beta(\ell,n)$ is a collection of arbirary real constants (note that $\beta(0,n)=0$). Thus finally,
\begin{equation}
\frac{1}{r_1}\,h(r_1,\theta_1;r_2,\theta_2)= i\sum_{\ell,n}
\beta(\ell,n)\,\varphi_{\ell,n}(r_1,\theta_1)\,\varphi_{-\ell,n}(r_2,\theta_2).
\end{equation}

Having identified the triplet of functions $(f,g,h)$ a direct substitution into (\ref{eq:Q2}) determines
the explicit form for the general generator of all global (and dynamical) symmetries of the free field $\phi(t,r,\theta)$
in terms of its creation and annihilation operators. A patient but straightforward calculation then establishes that,
once brought into normal ordered form in the case of the quantum operator,
\begin{eqnarray}
Q(t) &=& \sum_{\ell,n}\alpha_+(\ell,n)\,(-1)^\ell\,\omega_{\ell,n}\,a^\dagger(\ell,n)a(-\ell,n)\ + \ \nonumber \\
&+&\sum_{\ell,n}\left(\alpha_-(\ell,n)\,\omega_{\ell,n} + \beta(\ell,n)\right)
(-1)^\ell\,a^\dagger(\ell,n) a(\ell,n).
\end{eqnarray}
Note that this operator is such that $Q^\dagger(t)=Q(t)$, as it should. Furthermore, since
one obviously has $[Q(t),\hat{H}]=0$ (for which the degeneracy property
$\omega_{-\ell,n}=\omega_{\ell,n}$ is crucial), even though our approach accounted for the possibility that
the generator of the general global symmetry could possess an explicit time dependence as a function
defined over phase space, it turns out that in fact the conserved quantity $Q(t)$ does not possess any explicit
time dependence.

All the possible global symmetries of the free field $\phi(t,r,\theta)$ have thus been identified,
of which the complete set of conserved and time independent Noether generators is composed of
the following bilinear operators in the Fock algebra generators,
\begin{equation}
N_{\ell,n}=a^\dagger(\ell,n)\,a(\ell,n),\qquad
Q_{\ell,n}=a^\dagger(\ell,n)\,a(-\ell,n),\qquad \ell\in\mathbb{Z},\quad n=1,2,3,\cdots,
\end{equation}
such that
\begin{equation}
N^\dagger_{\ell,n}=N_{\ell,n},\qquad
Q^\dagger_{\ell,n}=Q_{-\ell,n},\qquad
Q_{0,n}=N_{0,n},
\end{equation}
and with as constant group parameters essentially the linearly independent and arbitrary coefficients
$\alpha_+(\ell,n)\in\mathbb{C}$ and
$\alpha_-(\ell,n), \beta(\ell,n)\in\mathbb{R}$ such that
$\alpha^*_+(\ell,n)=\alpha_+(-\ell,n)$, $\alpha_-(-\ell,n)=\alpha_-(\ell,n)$, and $\beta(-\ell,n)=-\beta(\ell,n)$
(thus $\beta(0,n)=0$).

\section{The Complete Global Symmetry Group}
\label{Sect4}

The algebra of all global symmetries of the 2+1 dimensional free scalar field theory on the disk $D(R)$
is thus spanned by the operators $N_{\ell,n}$ and $Q_{\ell,n}$, respectively the number operator and the
angular-momentum flipping operator for each of the $(\ell,n)$ modes of the stationary waves of the field
and its conjugate momentum inside the disk. It should be clear these transformations do indeed transform a solution
to the Klein-Gordon equation into another solution to the same Klein-Gordon equation, and this without changing
the total energy of that field configuration since $\omega_{-\ell,n}=\omega_{\ell,n}$.
Furthermore in particular, note that the conserved total energy and angular-momentum of the
system are part of that large dynamical symmetry, which is much larger than the finite dimensional global symmetry
group of the underlying spacetime geometry with the topology of $\mathbb{R}\times D(R)$, namely
time translations and disk rotations of which, when acting on the field and its conjugate momentum
the generators are, respectively\footnote{On account of the completeness relation (\ref{eq:complete})
these two conserved Noether charges correspond to the choices $g(1;2)=\delta(r_1-r_2)\delta(\theta_1 - \theta_2/r_1$
and $h(1;2)=0$ in the case of $H$, and to the choices $g(1;2)=0$ and
$h(1;2)=\delta(r_1-r_2)\partial_{\theta_1}\delta(\theta_1-\theta_2)$ in the case of $L$, thereby being also
spatially local bilinear quantities in phase space.},
\begin{equation}
\hat{H}=\sum_{\ell,n}\omega_{\ell,n}\,N_{\ell,n},\qquad
\hat{L}=\sum_{\ell,n}\,\ell\,N_{\ell,n}.
\end{equation}

The commutator algebra generated by all these conserved Noether charges of course closes,
and is given by,
\begin{eqnarray}
\left[N_{\ell_1,n_1},N_{\ell_2,n_2}\right] &=& 0, \nonumber \\
\left[N_{\ell_1,n_1},Q_{\ell_2,n_2}\right] &=& \delta_{\ell_1,\ell_2}\delta_{n_1,n_2}\,Q_{\ell_1,n_1}\,-\,
\delta_{\ell_1,-\ell_2}\delta_{n_1,n_2}\,Q_{-\ell_1,n_1}, \\
\left[Q_{\ell_1,n_1},Q_{\ell_2,n_2}\right] &=&
\delta_{\ell_1,-\ell_2}\delta_{n_1,n_2}\left(N_{\ell_1,n_1} - N_{-\ell_1,n_1}\right).
\end{eqnarray}
In particular it also follows that,
\begin{equation}
\left[\hat{H},N_{\ell,n}\right]=0,\qquad
\left[\hat{L},N_{\ell,n}\right]=0,
\end{equation}
\begin{equation}
\left[\hat{H},Q_{\ell,n}\right]=0,\qquad
\left[\hat{L},Q_{\ell,n}\right]=2\ell\,Q_{\ell,n},
\end{equation}
as it should of course, while furthermore,
\begin{eqnarray}
\left[N_{\ell_1,n_1},a(\ell_2,n_2)\right]=-\delta_{\ell_1,\ell_2}\delta_{n_1,n_2}\,a(\ell_1,n_1) &,&
\left[N_{\ell_1,n_1},a^\dagger(\ell_2,n_2)\right]=+\delta_{\ell_1,\ell_2}\delta_{n_1,n_2}\,a^\dagger(\ell_1,n_1),
\nonumber \\
&& \\
\left[Q_{\ell_1,n_1},a(\ell_2,n_2)\right]=-\delta_{\ell_1,\ell_2}\delta_{n_1,n_2}\,a(-\ell_1,n_1) &,&
\left[Q_{\ell_1,n_1},a^\dagger(\ell_2,n_2)\right]=+\delta_{\ell_1,-\ell_2}\delta_{n_1,n_2}\,a^\dagger(\ell_1,n_1).
\nonumber
\end{eqnarray}

In order to identify the symmetry group associated to the above algebra generated by $N_{\ell,n}$ and $Q_{\ell,n}$
let us take as a clue the expressions for the total Hamiltonian and angular-momentum operators
in (\ref{eq:Hsu2}) and (\ref{eq:Lsu2}). Clearly quantum sectors with different values for $n=1,2,3,\cdots$
are all decoupled form one another, and so are the sectors with different values of $|\ell |$ for $\ell\in\mathbb{Z}$.
However given a fixed value for $n$, the sectors with opposite values of $\ell\ne 0$ and $-\ell$ are coupled
to one another through the action of the global symmetries generated by $N_{\ell,n}$ and $Q_{\ell,n}$.

Let us thus first consider the sector with $\ell=0$ and a given value for $n=n_0$. Since such a sector does
not contribute to the total angular-momentum $\hat{L}$, is left invariant by the sole symmetry generator
$N_{0,n_0}=Q_{0,n_0}=a^\dagger(0,n_0)a(0,n_0)$ and contributes to the total energy as
$\hat{H}(0,n_0)=\omega_{0,n_0}N_{0,n_0}$, any such sector is equivalent to that of an ordinary one-dimensional
harmonic oscillator of angular frequency $\omega_{0,n_0}$, of which the global symmetry is the
U(1)$_{0,n_0}$ phase symmetry generated by the number operator $N_{0,n_0}$. Hence the complete
global symmetry of all $\ell=0$ sectors of the free scalar field in the disk $D(R)$
is $\bigotimes_{n_0=1}^\infty U(1)_{0,n_0}$.

Consider now a specific non vanishing and positive value of $\ell=1,2,3,\cdots$
as well as a given value for $n=n_\ell$. Let us then adapt the notations as follows,
in order to emphasize the analogy to be highlighted,
\begin{equation}
a_+\equiv a(\ell,n_\ell),\qquad
a_-\equiv a(-\ell,n_\ell),\qquad
a^\dagger_+\equiv a^\dagger(\ell,n_\ell),\qquad
a^\dagger_-\equiv a^\dagger(-\ell,n_\ell),
\end{equation}
and consider the following combinations of the symmetry generators for the sectors $(\ell,n_\ell)$ and $(-\ell,n_\ell)$,
\begin{eqnarray}
T_0 &=& T_0(\ell,n_\ell)\equiv
\frac{1}{2}\left(N_{\ell,n_\ell}+N_{-\ell,n_\ell}\right)=\frac{1}{2}\left(a^\dagger_+ a_+ + a^\dagger_- a_-\right),
\nonumber \\
T_3 &=& T_3(\ell,n_\ell)\equiv
\frac{1}{2}\left(N_{\ell,n_\ell}-N_{-\ell,n_\ell}\right)=\frac{1}{2}\left(a^\dagger_+ a_+ - a^\dagger_- a_-\right), \\
T_+&=& T_+(\ell,n_\ell)\equiv Q_{\ell,n_\ell}=a^\dagger_+ a_- \equiv T_1 + i T_2, \nonumber \\
T_-&=& T_-(\ell,n_\ell)\equiv Q_{-\ell,n_\ell}=a^\dagger_- a_+ \equiv T_1 - i T_2.
\end{eqnarray}
Obviously $2\omega_{\ell,n}T_0$ is the total contribution of the sectors $(\ell,n)$ and $(-\ell,n)$ to the
total energy of the system, while $2\ell T_3$ is that to its total angular-momentum. Once again $T_0$
generates a global U(1)$_{\ell,n_\ell}$ phase symmetry for these two sectors of the system.
However one has furthermore
\begin{equation}
[T_3,T_+]=+T_+,\qquad
[T_3,T_-]=-T_-,\qquad
[T_+,T_-]=2T_3,
\end{equation}
or equivalently,
\begin{equation}
[T_i,T_j]=i\epsilon_{ijk}\,T_k,\qquad \epsilon_{123}=+1,\qquad i,j,k=1,2,3,
\end{equation}
in which one recognises the SU(2) Lie algebra. Indeed the Fock algebras $(a_\pm,a^\dagger_\pm)$
are precisely those of the two dimensional spherically symmetric harmonic oscillator in the Euclidean plane
(in the helicity basis), which is well known to possess a SU(2) dynamical symmetry with the above SU(2) algebra.
Thus the sectors $(\ell,n_\ell)$ and $(-\ell,n_\ell)$ with $\ell\ne 0$ of the free scalar field in the disk $D(R)$
possess a global and dynamical SU(2)$_{\ell,n_\ell}$ symmetry, which is not spatially local in phase space.

Consequently one comes to the final conclusion that the complete global symmetry group of the free scalar field
in the disk $D(R)$ -- which for most of it is a dynamical and spatially non local symmetry group -- is identified as
the following infinite countable symmetry group
\begin{equation}
\bigotimes_{n_0=1}^\infty U(1)_{0,n_0}\bigotimes_{\ell=1}^\infty\bigotimes_{n_\ell=1}^\infty U(2)_{\ell,n_\ell}.
\end{equation}
The finite dimensional symmetry group of the underlying spacetime geometry, namely the direct product
of global time translations and space rotations, is a specific subgroup of the above, with the following two abelian
generators which are also spatially local in phase space,
\begin{equation}
\hat{H}=\sum_{n_0=1}^\infty\omega_{0,n_0}N_{0,n_0}
+\sum_{\ell=1}^\infty\sum_{n_\ell=1}^\infty\,2\omega_{\ell,n}\,T_0(\ell,n_\ell),\qquad
\hat{L}=\sum_{\ell=1}^\infty\sum_{n_\ell=1}^\infty\,2\ell\,T_3(\ell,n_\ell).
\end{equation}

To make the above points somewhat more explicit, let us also consider now the finite symmetry transformations
generated by each of the generators $N_{\ell,n}$, $Q_{\ell,n}$ and $Q^\dagger_{\ell,n}=Q_{-\ell,n}$
to check that indeed they map solutions into other solutions to the equations of motion, by redefining
and mixing the mode amplitudes $a(\ell,n)$ and $a^\dagger(\ell,n)$ and thus the latter's contributions
to the field and its conjugate momentum, $\phi(t,r,\theta)$ and $\pi(t,r,\theta)$, in the Heisenberg picture.

Beginning with the Hermitian number operators $N_{\ell,n}=N^\dagger_{\ell,n}$, let us consider a specific
but otherwise arbitrary choice of values $\ell_0\in\mathbb{Z}$ and $n_0\in\mathbb{N}^+$ and the number operator
$N_{\ell_0,n_0}=a^\dagger(\ell_0,n_0)a(\ell_0,n_0)$ corresponding to the mode $(\ell_0,n_0)$ of the field.
Since the only sector $(\ell,n)$ and its Fock algebra which does not commute with $N_{\ell_0,n_0}$
is that of the mode $(\ell_0,n_0)$, all finite and global symmetries generated by that number operator
through the action of the unitary operator
\begin{equation}
U(\alpha)\equiv e^{i\alpha N_{\ell_0,n_0}},\qquad \alpha\in\mathbb{R},
\end{equation}
leave invariant all sectors $(\ell,n)\ne(\ell_0,n_0)$, and only act on the single sector $(\ell_0,n_0)$.
In particular given that
\begin{equation}
\left[N_{\ell_0,n_0},a(\ell_0,n_0)\right]=-a(\ell_0,n_0),\qquad
\left[N_{\ell_0,n_0},a^\dagger(\ell_0,n_0)\right]=+a^\dagger(\ell_0,n_0),
\end{equation}
one readily establishes that
\begin{eqnarray}
\tilde{a}(\ell_0,n_0) &\equiv& U(\alpha)\,a(\ell_0,n_0)\,U^\dagger(\alpha)=e^{-i\alpha}\,a(\ell_0,n_0), \nonumber \\
\tilde{a}^\dagger(\ell_0,n_0) &\equiv& U(\alpha)\,a^\dagger(\ell_0,n_0)\,U^\dagger(\alpha)
=e^{i\alpha}\,a^\dagger(\ell_0,n_0).
\end{eqnarray}
Note that these unitary transformations leave invariant the Fock algebra of the creation and annihilation
operators of the sector $(\ell_0,n_0)$, as well as its Fock vacuum. When acting on the phase space fields
as $U(\alpha) A\, U^\dagger(\alpha)$ where $A$ is any operator, clearly these transformations only lead
to a phase redefinition of the mode amplitudes $a(\ell_0,n_0)$ and $a^\dagger(\ell_0,n_0)$,
thereby indeed mapping any given solution to some other
solution of the same Hamiltonian equations of motion of the Klein-Gordon scalar field in the disk.

Turning now to the angular-momentum flipping operators $Q_{\ell,n}$, since $Q_{0,n}=N_{0,n}$
let us consider a specific but otherwise arbitrary choice of values $\ell_0\in\mathbb{N}^+$
and $n_0\in\mathbb{N}^+$ and the operators $Q_{\ell_0,n_0}=a^\dagger(\ell_0,n_0)a(-\ell_0,n_0)$
and $Q_{-\ell_0,n_0}=Q^\dagger_{\ell_0,n_0}=a^\dagger(-\ell_0,n_0)a(\ell_0,n_0)$
corresponding to the modes $(\ell_0,n_0)$ and $(-\ell_0,n_0)$ of the field. In this case all sectors
except for these two are left invariant under any of the finite global symmetry transformations
generated by these two operators. Let us recall that these transformations are indeed symmetries
because of the degeneracy $\omega_{-\ell_0,n_0}=\omega_{\ell_0,n_0}$.

In order to work with Hermitian operators, let us introduce the operators,
\begin{equation}
T_1(\ell_0,n_0)=\frac{1}{2}\left(Q_{\ell_0,n_0}+Q^\dagger_{\ell_0,n_0}\right),\qquad
T_2(\ell_0,n_0)=-\frac{1}{2}i\left(Q_{\ell_0,n_0} - Q^\dagger_{\ell_0,n_0}\right),
\end{equation}
such that $T^\dagger_1(\ell_0,n_0)=T_1(\ell_0,n_0)$ and $T^\dagger_2(\ell_0,n_0)=T_2(\ell_0,n_0)$.
Given that one has in this case,
\begin{eqnarray}
\left[T_1(\ell_0,n_0),a(\pm\ell_0,n_0)\right]=-\frac{1}{2}a(\mp\ell_0,n_0) &,&
\left[T_2(\ell_0,n_0),a(\pm\ell_0,n_0)\right]=\pm\frac{1}{2} i a(\mp\ell_0,n_0) , \nonumber \\
\left[T_1(\ell_0,n_0),a^\dagger(\pm\ell_0,n_0)\right]=+\frac{1}{2} a^\dagger(\mp\ell_0,n_0) &,&
\left[T_2(\ell_0,n_0),a^\dagger(\pm\ell_0,n_0)\right]=\pm\frac{1}{2} i a^\dagger(\mp\ell_0,n_0),
 \end{eqnarray}
 the finite global and unitary symmetry transformations
 \begin{equation}
 U(\alpha_1)\equiv e^{i\alpha_1 T_1(\ell_0,n_0)},\qquad
 U(\alpha_2)\equiv e^{i\alpha_2 T_2(\ell_0,n_0)},\qquad
 \alpha_1,\alpha_2\in\mathbb{R},
 \end{equation}
 generate the following redefinitions and mixings of the mode amplitudes of the sectors $(\ell_0,n_0)$
 and $(-\ell_0,n_0)$, which are hence indeed once again genuine symmetry transformations mapping solutions
 into other solutions to the dynamical equations of the field, namely first for transformations generated by $T_1$,
 \begin{eqnarray}
 \tilde{a}(\pm\ell_0,n_0) &\equiv& U(\alpha_1)\,a(\pm\ell_0,n_0)\,U^\dagger(\alpha_1)
 =\cos\frac{\alpha_1}{2}\,a(\pm\ell_0,n_0)\,-\,i\sin\frac{\alpha_1}{2}\,a(\mp\ell_0,n_0), \nonumber \\
 \tilde{a}^\dagger(\pm\ell_0,n_0) &\equiv& U(\alpha_1)\,a^\dagger(\pm\ell_0,n_0)\,U^\dagger(\alpha_1)
 =\cos\frac{\alpha_1}{2}\,a^\dagger(\pm\ell_0,n_0)\,+\,i\sin\frac{\alpha_1}{2}\,a^\dagger(\mp\ell_0,n_0),
 \end{eqnarray}
 as well as for transformations generated by $T_2$,
  \begin{eqnarray}
 \tilde{a}(\pm\ell_0,n_0) &\equiv& U(\alpha_2)\,a(\pm\ell_0,n_0)\,U^\dagger(\alpha_2)
 =\cos\frac{\alpha_2}{2}\,a(\pm\ell_0,n_0)\,\mp\,\sin\frac{\alpha_2}{2}\,a(\mp\ell_0,n_0), \nonumber \\
 \tilde{a}^\dagger(\pm\ell_0,n_0) &\equiv& U(\alpha_2)\,a^\dagger(\pm\ell_0,n_0)\,U^\dagger(\alpha_2)
 =\cos\frac{\alpha_2}{2}\,a^\dagger(\pm\ell_0,n_0)\,\mp\,\sin\frac{\alpha_2}{2}\,a^\dagger(\mp\ell_0,n_0).
 \end{eqnarray}
 Note again that these unitary transformations leave invariant the Fock algebras of both sectors
 $(\ell_0,n_0)$ and $(-\ell_0,n_0)$, as well as the corresponding Fock vacua.

\section{Conclusions}
\label{SectConclusions}

In view of the double degeneracy in the energy spectrum under the exchange $\ell\leftrightarrow -\ell$ for $\ell\ne 0$
as made explicit in (\ref{eq:Hsu2}) and (\ref{eq:Lsu2}), and given the
hindsight gained through the present analysis, it would appear rather obvious that indeed the
system possesses as finite global (and dynamical) symmetries the group identified above, namely
\begin{equation}
\bigotimes_{n_0=1}^\infty U(1)_{0,n_0}\bigotimes_{\ell=1}^\infty\bigotimes_{n_\ell=1}^\infty U(2)_{\ell,n_\ell}.
\end{equation}
However that there do not exist any further global symmetries is less obvious, a conclusion which
requires an approach such as the one having been used herein and based on the rationale outlined
in the Introduction.
This is not to say that there do not exist other conserved quantities for the system. Rather it means
that there do not exist conserved quantities other than $N_{\ell,n}$ and $Q_{\ell,n}$ that would
also be generators of additional continuous global symmetries of the system.

To make this point more explicit, given specific but otherwise arbitrary values for $\ell_1$ and $\ell_2$
and for $n_1$ and $n_2$, and such that $|\ell_1|\ne|\ell_2|$, consider the following operators
bilinear in the creation and annihilation operators for the mode sectors $(\ell_1,n_1)$ and $(\ell_2,n_2)$,
\begin{eqnarray}
Q(\ell_1,n_1;\ell_2,n_2;t) &=& e^{i(\omega_{\ell_2,n_2}-\omega_{\ell_1,n_1})t}\,a^\dagger(\ell_1,n_1)\,a(\ell_2,n_2),
\nonumber \\
Q^\dagger(\ell_1,n_2;\ell_2,n_2;t) &=& e^{i(\omega_{\ell_1,n_1}-\omega_{\ell_2,n_2})t}\,
a^\dagger(\ell_2,n_2)\, a(\ell_1,n_1)=Q(\ell_2,n_2;\ell_1,n_1;t),
\end{eqnarray}
in which of course then $\omega_{\ell_1,n_1}\ne\omega_{\ell_2,n_2}$. Even though they do not commute
with the Hamiltonian, $\hat{H}$, of the system these operators are conserved quantities because of their specific
explicit time dependence such that their Heisenberg equation of motion reads,
\begin{equation}
i\frac{dQ(\ell_1,n_1;\ell_2,n_2;t)}{dt}=i\frac{\partial Q(\ell_1,n_1;\ell_2,n_2;t)}{\partial t}
\ +\ \left[Q(\ell_1,n_1;\ell_2,n_2;t),H\right]=0.
\end{equation} 
Consider now the two associated conserved Hermitian operators
\begin{eqnarray}
Q_+(t)\equiv \frac{1}{2}\left(Q(\ell_1,n_1;\ell_2,n_2;t) + Q^\dagger(\ell_1,n_1;\ell_2,n_2;t)\right), \nonumber \\
Q_-(t)\equiv -\frac{1}{2}i\left(Q(\ell_1,n_1;\ell_2,n_2;t) - Q^\dagger(\ell_1,n_1;\ell_2,n_2;t)\right).
\end{eqnarray}
Using the shorthand notations $\omega_1\equiv\omega_{\ell_1,n_1}$, $\omega_2\equiv\omega_{\ell_2,n_2}$,
$a_1\equiv a(\ell_1,n_1)$, $a_2\equiv a(\ell_2,n_2)$, $a^\dagger_1\equiv a^\dagger(\ell_1,n_1)$
and $a^\dagger_2\equiv a^\dagger(\ell_2,n_2)$, one then has the relations,
\begin{eqnarray}
\left[Q_+(t),a_1\right] = e^{i(\omega_2 -\omega_1)t}\,\frac{-1}{2}\,a_2 &,& 
\left[Q_-(t),a_1\right] = e^{i(\omega_2 -\omega_1)t}\,\frac{i}{2}\,a_2, \nonumber \\
\left[Q_+(t),a_2\right] = e^{i(\omega_1 -\omega_2)t}\,\frac{-1}{2}\,a_1&,&
\left[Q_-(t),a_2\right] = e^{i(\omega_1 -\omega_2)t}\,\frac{-i}{2}\,a_1, \nonumber \\
\left[Q_+(t),a^\dagger_1\right] = e^{i(\omega_1 -\omega_2)t}\,\frac{1}{2}\,a^\dagger_2 &,&
\left[Q_-(t),a^\dagger_1\right] = e^{i(\omega_1 -\omega_2)t}\,\frac{i}{2}\,a^\dagger_2, \\
\left[Q_+(t),a^\dagger_2\right] = e^{i(\omega_2 -\omega_1)t}\,\frac{1}{2}\,a^\dagger_1 &,&
\left[Q_-(t),a^\dagger_2\right] = e^{i(\omega_2 -\omega_1)t}\,\frac{-i}{2}\,a^\dagger_1.
\end{eqnarray}
For finite transformations generated by $Q_+(t)$ and $Q_-(t)$ it then follows that, given $\alpha\in\mathbb{R}$,
\begin{eqnarray}
e^{i\alpha Q_+(t)}\,a_1\,e^{-i\alpha Q_+(t)} &=& 
\cos\frac{\alpha}{2}\,a_1\ -\ i\sin\frac{\alpha}{2}\,e^{i(\omega_2 - \omega_1)t}\,a_2, \nonumber \\
e^{i\alpha Q_+(t)}\,a_2\,e^{-i\alpha Q_+(t)} &=& 
\cos\frac{\alpha}{2}\,a_2\ -\ i\sin\frac{\alpha}{2}\,e^{i(\omega_1 - \omega_2)t}\,a_1, \nonumber \\
e^{i\alpha Q_+(t)}\,a^\dagger_1\,e^{-i\alpha Q_+(t)} &=& 
\cos\frac{\alpha}{2}\,a^\dagger_1\ +\ i\sin\frac{\alpha}{2}\,e^{i(\omega_1 - \omega_2)t}\,a^\dagger_2,  \\
e^{i\alpha Q_+(t)}\,a^\dagger_2\,e^{-i\alpha Q_+(t)} &=& 
\cos\frac{\alpha}{2}\,a^\dagger_2\ +\ i\sin\frac{\alpha}{2}\,e^{i(\omega_2 - \omega_1)t}\,a^\dagger_1, \nonumber \\
\end{eqnarray}
as well as,
\begin{eqnarray}
e^{i\alpha Q_-(t)}\,a_1\,e^{-i\alpha Q_-(t)} &=& 
\cos\frac{\alpha}{2}\,a_1\ -\ \sin\frac{\alpha}{2}\,e^{i(\omega_2 - \omega_1)t}\,a_2, \nonumber \\
e^{i\alpha Q_-(t)}\,a_2\,e^{-i\alpha Q_-(t)} &=& 
\cos\frac{\alpha}{2}\,a_2\ +\ \sin\frac{\alpha}{2}\,e^{i(\omega_1 - \omega_2)t}\,a_1, \nonumber \\
e^{i\alpha Q_-(t)}\,a^\dagger_1\,e^{-i\alpha Q_-(t)} &=& 
\cos\frac{\alpha}{2}\,a^\dagger_1\ -\ \sin\frac{\alpha}{2}\,e^{i(\omega_1 - \omega_2)t}\,a^\dagger_2,  \\
e^{i\alpha Q_-(t)}\,a^\dagger_2\,e^{-i\alpha Q_-(t)} &=& 
\cos\frac{\alpha}{2}\,a^\dagger_2\ +\ \sin\frac{\alpha}{2}\,e^{i(\omega_2 - \omega_1)t}\,a^\dagger_1. \nonumber \\
\end{eqnarray}
Note that these two classes of unitary transformations leave invariant the Fock algebras of the mode sectors
$(\ell_1,n_1)$ and $(\ell_2,n_2)$, as it should. However clearly they do not map solutions
to the Klein-Gordon equation into some other solutions to the same equation, and thus do not define symmetries
of the system. Even though conserved, the operators $Q_+(t)$ and $Q_-(t)$ are not generators of any global
symmetry of the system.

These considerations are indeed in perfect agreement with the conclusions of the detailed and general
analysis developed in the present paper, which established that all global (and dynamical) symmetries
of the system are generated by the conserved operators $N_{\ell,n}$ and $Q_{\ell,n}$ and only those bilinear
operators in the creation and annihilation operators of field standing modes.
By considering all possible bilinear quantities in the phase space degrees of freedom, even those
that are not spatially local, and by requiring the Hamiltonian action to be left invariant up to a total
time derivative under variations of the phase space degrees of freedom generated by all such spatially bilocal
phase space bilinear quantities, provides a systematic rationale for identifying all possible
global (and dynamical) symmetries of this system with linear equations of motion.

Relying on the understanding achieved through that approach in the present case of a free field theory
restricted to a bounded spatial domain, in a forthcoming work we plan to
apply the same rationale to a free scalar field theory defined this time over an unbounded
Minkowski spacetime, in order to also establish then the relation between all global (and dynamical)
symmetries of that system with its BMS symmetries.

\section*{Acknowledgements}

DBI acknowledges the support of an ``extraordinary" postdoctoral Fellowship of the
{\it Acad\'emie de Recherche et d'Enseignement Sup\'erieur} (ARES) of the Wallonia-Brussels
Federation of Belgium towards a six months stay at CP3 during which the present work was
completed. The work of JG is supported in part by the Institut Interuniversitaire des Sciences
Nucl\'eaires (IISN, Belgium).

\end{document}